%% file: ms.tex
\documentstyle[emulateapj,apjfonts,epsf]{article}

\def\lapp{\ifmmode\stackrel{<}{_{\sim}}\else$\stackrel{<}{_{\sim}}$\fi}
\def\gapp{\ifmmode\stackrel{>}{_{\sim}}\else$\stackrel{<}{_{\sim}}$\fi}

\begin{document}

\title{Three Binary Millisecond Pulsars in NGC~6266}

\author{A.~Possenti,\altaffilmark{1,2}
N.~D'Amico,\altaffilmark{2,3}
R.~N.~Manchester,\altaffilmark{4} 
F.~Camilo,\altaffilmark{5}
A.~G.~Lyne,\altaffilmark{6}
J.~Sarkissian,\altaffilmark{7}
and A.~Corongiu\altaffilmark{2,3}}
\medskip

\affil{\altaffilmark{1}Osservatorio Astronomico di Bologna,
Via Ranzani 1, 40127 Bologna, Italy}
\affil{\altaffilmark{2}Osservatorio Astronomico di Cagliari,
Loc. Poggio dei Pini, Strada 54, 09012 Capoterra (CA), Italy}
\affil{\altaffilmark{3}Dipartimento di Fisica, Universit\`a di Cagliari,
Strada Provinciale Monserrato--Sestu, km 0.700, I--09042 Monserrato, Italy}
\affil{\altaffilmark{4}Australia Telescope National Facility,
CSIRO, PO Box 76, Epping, NSW 1710, Australia}
\affil{\altaffilmark{5}Columbia Astrophysics Laboratory, 
Columbia University, 550 West 120th Street, New York, NY~10027}
\affil{\altaffilmark{6}University of Manchester, Jodrell Bank
Observatory, Macclesfield, Cheshire SK11~9DL, UK}
\affil{\altaffilmark{7}Australia Telescope National Facility,
CSIRO, Parkes Observatory, PO Box 276, Parkes, NSW 2870, Australia}

\bigskip

\begin{abstract}
We present rotational and astrometric parameters of three millisecond
pulsars located near the center of the globular cluster NGC~6266 (M62)
resulting from timing observations with the Parkes radio
telescope.  Their accelerations toward the cluster center yield values
of the cluster central density and mass-to-light ratio consistent with
those derived from optical data.  The three pulsars are in binary
systems. One (spin period $P=$5.24 ms) is in a 3.5-day orbit
around a companion of minimum mass $0.2~{\rm M_\odot}$. The other two 
millisecond pulsars ($P=$3.59 ms and 3.81 ms) have shorter
orbital periods (3.4~h and 5.0~h) and lighter companions (minimum mass
$ 0.12~{\rm M_\odot}$ and $0.07~{\rm M_\odot}$ respectively).  The
pulsar in the closest system is the fifth member of an emerging class
of millisecond pulsars displaying irregular radio eclipses and having
a relatively massive companion.  This system is a good candidate for
optical identification of the companion star. The lack of known
isolated pulsars in NGC~6266 is also discussed.
\end{abstract}


\section{Introduction}
\label{Intro}

Recycled pulsars are old neutron stars revived through transfer of
matter and angular momentum from a mass-donor companion in a binary
system (e.g. Alpar et al. 1982\nocite{acr+82}; Smarr \& Blandford
1976\nocite{sb76}; Bhattacharya \& van den Heuvel 1991\nocite{bv91}).
They are point-like objects and can be considered as test masses for
probing gravitational effects. Most of them are also extremely stable
clocks, allowing for accurate measurements of their rotational
parameters, position and apparent motion in the sky.  Because of these
characteristics, recycled pulsars found in globular clusters (GCs)
have proven to be valuable tools for studying the GC potential well
(e.g. Phinney 1992\nocite{phi92}; Camilo et al. 2000\nocite{clf+00};
D'Amico et al. 2002\nocite{dpf+02}), the dynamical interactions in GC
cores (e.g. Phinney \& Sigurdsson 1991\nocite{ps91}; Colpi, Possenti
\& Gualandris 2002\nocite{cpg02}) and neutron star retention in GCs
(e.g. Rappaport et al. 2001\nocite{r++01}). In the case of 47 Tucanae
they allowed also the first detection of gas in a GC
(Freire et al. 2001\nocite{fklcmd01}).

Globular clusters are a fertile environment for the formation of
recycled pulsars: besides evolution from a primordial system, exchange
interactions in the ultra-dense core of the cluster favor the
formation of various types of binary systems suitable for spinning up the
neutron stars they host (Davies \& Hansen 1998\nocite{dh98}).  Because
of this, about 60\% of all known millisecond pulsars (MSPs) are
in GCs.  Unfortunately, pulsars in GCs are elusive sources since they
are often distant and in close binary systems.  Their large
distances make their flux density typically very small and their
signals strongly distorted by propagation through the dispersive
interstellar medium. In addition, they frequently are members of close
binary systems, causing large changes in the apparent spin
period and sometimes periodic eclipsing of the radio signal.

The Parkes Globular Cluster (PKSGC) survey is a search for pulsars in
the system of southern GCs using the Parkes 64-m radio telescope which
commenced in 2000.  It exploits the high sensitivity of the central
beam of the Parkes multibeam receiver (Staveley-Smith et
al. 1996\nocite{ss+96}), the efficiency of a modern data acquisition
system (e.g., Manchester et al. 2001\nocite{m++01}, D'Amico et
al. 2001a\nocite{dlm+01}) and the high resolution of a new filterbank
designed and assembled at Jodrell Bank and Bologna, with the aim of
improving the capability for probing distant clusters. Time series
data are analyzed with a modern algorithm for the incoherent search of
periodicities over a range of dispersion measures (DMs) and
accelerations resulting from orbital motion.

This project has already resulted in the discovery of 12
millisecond pulsars in six globular clusters which had no previously
associated pulsar (D'Amico et al. 2001a\nocite{dlm+01}; D'Amico et
al. 2001b\nocite{dpm+01b}; Possenti et al. 2001\nocite{pdm+01};
D'Amico et al. 2002\nocite{dpf+02}).  These detections reversed the
declining trend in discoveries of additional clusters hosting these
objects; in the seven years from 1987 (when the first pulsar in a GC,
B1821$-$24 in M28, was discovered at Jodrell Bank by Lyne et
al. 1987\nocite{lbmkb87}) to 1994 (B1820-30A and B in NGC~6624: Biggs
et al. 1994\nocite{bbfglb94}) 13 globular clusters were shown to
contain at least one pulsar, whereas no new cluster joined the list in
the following six years. More recently, pulsars have been detected in
a further three GCs (Ransom 2003a\nocite{r+03}; Ransom
2003b\nocite{r03b}; Jacoby 2003\nocite{j03}), bringing the current
total to 73 pulsars in 22 clusters.\footnote{The association of the
long-period pulsar B1718$-$19 with the cluster NGC~6342,
questioned by some, is included in this list.}

This paper discusses results from the PKSGC survey of NGC~6266
(M62). The discovery of the first pulsar in this cluster, PSR
J1701$-$3006A, was presented by D'Amico et al. (2001a\nocite{dlm+01}).
A preliminary announcement of the discovery of two more MSPs was also
made by D'Amico et al. (2001b\nocite{dpm+01b}), while three further
millisecond pulsars were later detected at the Green Bank
Telescope (Jacoby et al. 2002\nocite{jcb+02}). Here we report details
of the discovery of the second and third pulsars, PSRs J1701$-$3006B
and J1701$-$3006C, both members of short-period binary systems,
and discuss timing results obtained over a 3-yr interval for all
three MSPs discovered at Parkes. Based on these results we
investigate the properties of the pulsars and of the host cluster.  We
particularly discuss the MSP in the tightest of the three
systems, which belongs to the rare class of eclipsing radio pulsars.

\section{Data collection and processing}
\label{Processing}

The PKSGC survey uses the dual polarization central beam of the 20-cm
multibeam receiver of the Parkes radio telescope.  The two channels
have a system temperature of $\sim 22$~K and a central frequency of
1390 MHz. A high-resolution filterbank system consisting of $512\times
0.5$ MHz adjacent channels per polarization is used to minimize
dispersion smearing, preserving significant sensitivity to a 3 ms
pulsar with dispersion measure up to 300 cm$^{-3}$pc. In this case,
the limiting sensitivity is $\sim 0.15$ mJy for a signal-to-noise ratio
(s/n) of 8 and a standard 2-h observation (assuming a typical duty
cycle of $\sim 20$\% and negligible scattering).  After adding the
outputs in polarization pairs, the resulting 512 data streams are each
high-pass-filtered, integrated and 1-bit digitized every 125
$\mu$s. Each observation typically produces $2-4$ Gbytes of data; a
cluster of 10 Alpha-500MHz CPUs at the Astronomical Observatory of
Bologna has been used for offline processing.

The processing pipeline first splits each data stream into
non-overlapping segments of 1050, 2100, 4200 or 8400 s, which are
processed separately. When no pulsar is known in a GC (and the DM is 
therefore unknown) the data are dedispersed over a wide range of $\sim
500-1000$ trial DMs, spanning the interval $(1.0 \pm 0.4)$~DM$_{\rm
exp}$, where DM$_{\rm exp}$ is the DM expected for the cluster
according to a model for the Galactic distribution of the ionized gas
(Taylor \& Cordes 1993\nocite{tc93}).  Each dedispersed series is then
transformed using a Fast Fourier Transform, and the resulting spectra
are searched for significant peaks. The process is repeated for
spectra obtained from summing 2, 4, 8 and 16 harmonics. This produces
a large number of candidate periods above a threshold. The time-domain
data are then folded in sub-integrations at each of these periods in
turn and searched for both a linear and a parabolic shift in pulse
phase.  A linear shift corresponds to a correction in the candidate
period, whereas a parabolic correction is a signature of acceleration
of the pulsar due to its orbital motion. Parameters for final pulse
profiles with significant s/n are displayed for visual inspection.
This processing scheme led to the discovery of the first pulsar in
NGC~6266 (D'Amico et al. 2001a\nocite{dlm+01}).

For an MSP in a very close orbit and with relatively high minimum
companion mass, the acceleration may undergo significant variations
during an observation. As a consequence, weak sources can be missed at
the confirmation stage if a constant acceleration term is applied to
the data.  Hence code has been developed at Bologna for searching both
the acceleration and the derivative of the acceleration in the
sub-integration arrays of interesting candidates. Spanning a cubic
space, the code searches also for the period in a small interval of values
around the nominal candidate period. Using this code we were able to
confirm two more binary millisecond pulsars in NGC~6266.

Once a pulsar is detected and confirmed in a cluster, the data are
reprocessed with dedispersion at the DM value of the newly discovered
pulsar. The resulting time series is then subject to a fully coherent
search for Doppler-distorted signals over a large range of
acceleration values.  Applying this extremely CPU-intensive procedure
to NGC~6266, exploring accelerations in the interval $|a|<35$ m
s$^{-2}$ for 35-min long segments (and $|a| < 17.5$ m s$^{-2}$
for 70-min long segments), resulted in no further discoveries.

{\vskip 0.5truecm 
\epsfxsize=8.5truecm 
\epsfysize=9.0truecm
\epsfbox{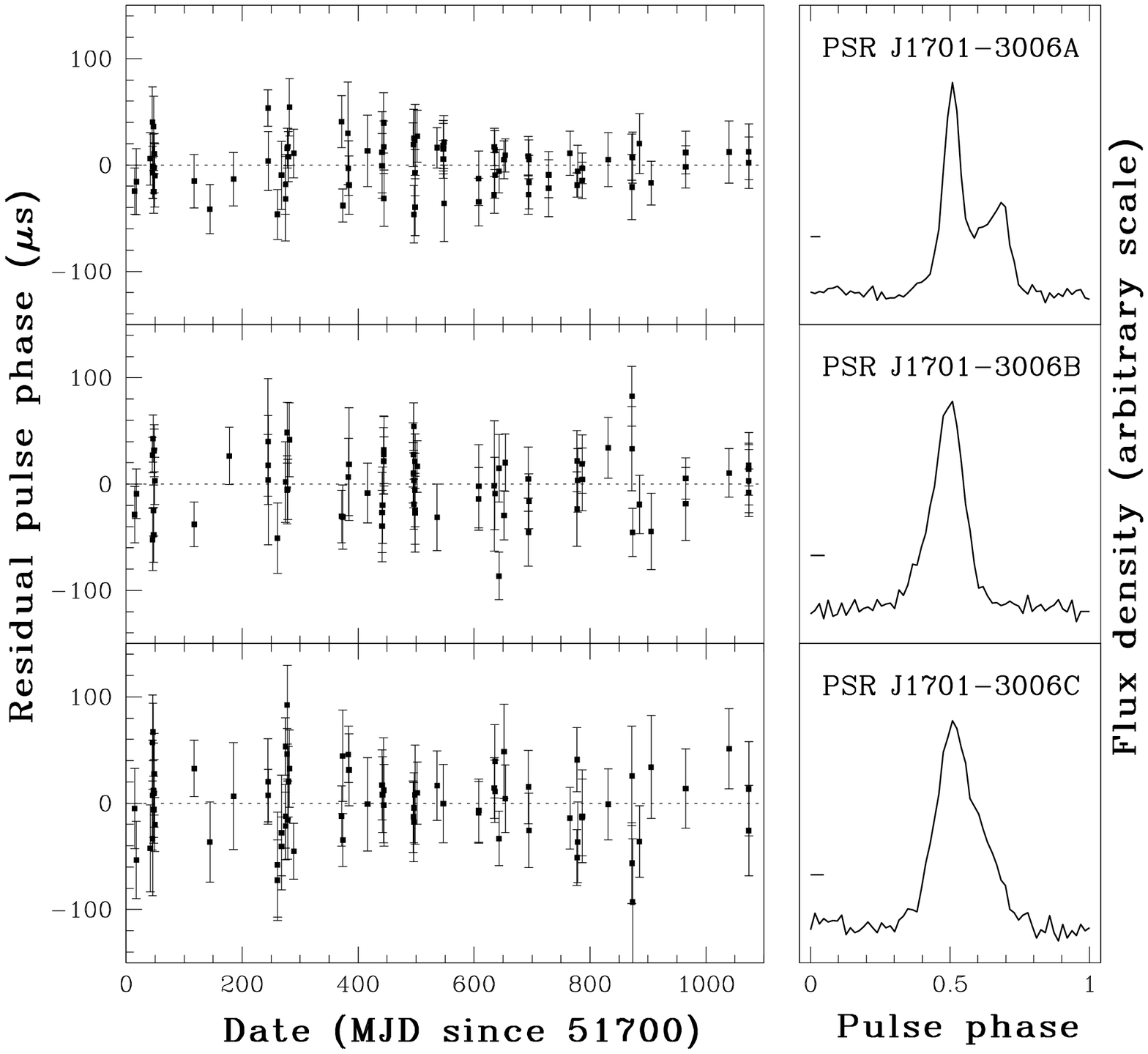} 
\figcaption[possenti1.eps]{\label{fig1}
{Post-fit timing residuals as a function of the Modified
Julian Day of observation ({\it left}) and integrated pulse profiles
at a central frequency of 1390 MHz ({\it right}) for the three millisecond
pulsars in NGC~6266 which are discussed in this paper. The short
horizontal line on the left side of each pulse profile represents the
time resolution of the integrated profile including DM smearing.}}
\vskip 0.5truecm}

Regular pulsar timing observations at the Parkes 64-m radio telescope
began shortly after the discovery of these pulsars, using the same
observing system as the search observations.  Timing observations,
typically of 30 to 60 minutes duration, are dedispersed and
synchronously folded at the predicted topocentric pulsar spin period
in an off-line process, forming pulse profiles every few minutes of
integration.  Topocentric pulse times of arrival (TOAs) are determined
by convolving a standard high s/n pulse template
with the observed pulse profiles and then analyzed using the program
{\sc tempo}\footnote{See
http://www.atnf.csiro.au/research/pulsar/timing/tempo.}. {\sc tempo}
converts the topocentric TOAs to solar-system barycentric TOAs at
infinite frequency (using the DE200 solar-system ephemeris, Standish
1982\nocite{sta82}) and then performs a multi-parameter fit to
determine the pulsar parameters.

Table~\ref{tab:pars} lists the timing parameters obtained for the
three pulsars, including precise positions. Values of the dispersion
measure (DM) were obtained for each pulsar by splitting the total
bandwidth into four adjacent 64-MHz wide sub-bands and computing the
differential delays. The available data do not yet allow a
constraining fit for the orbital eccentricity $e$ for any of the three
pulsars (see the footnotes to Table~\ref{tab:pars} for details of the
fitting procedure.) The mean flux densities at 1400 MHz ($S_{1400}$)
in Table~\ref{tab:pars} are average values, derived from the system
sensitivity estimate and the observed s/n.  In the case of
PSR~J1701$-$3006B, the quoted flux density refers only to epochs away
from the eclipse (see below).  As expected from the relatively high
DMs, interstellar scintillation does not significantly affect the
detectability of these sources; observed variations are within 30\% of
the nominal flux density reported in Table~\ref{tab:pars}.

{\vskip 0.5truecm 
\epsfxsize=8.5truecm 
\epsfysize=9.0truecm
\epsfbox{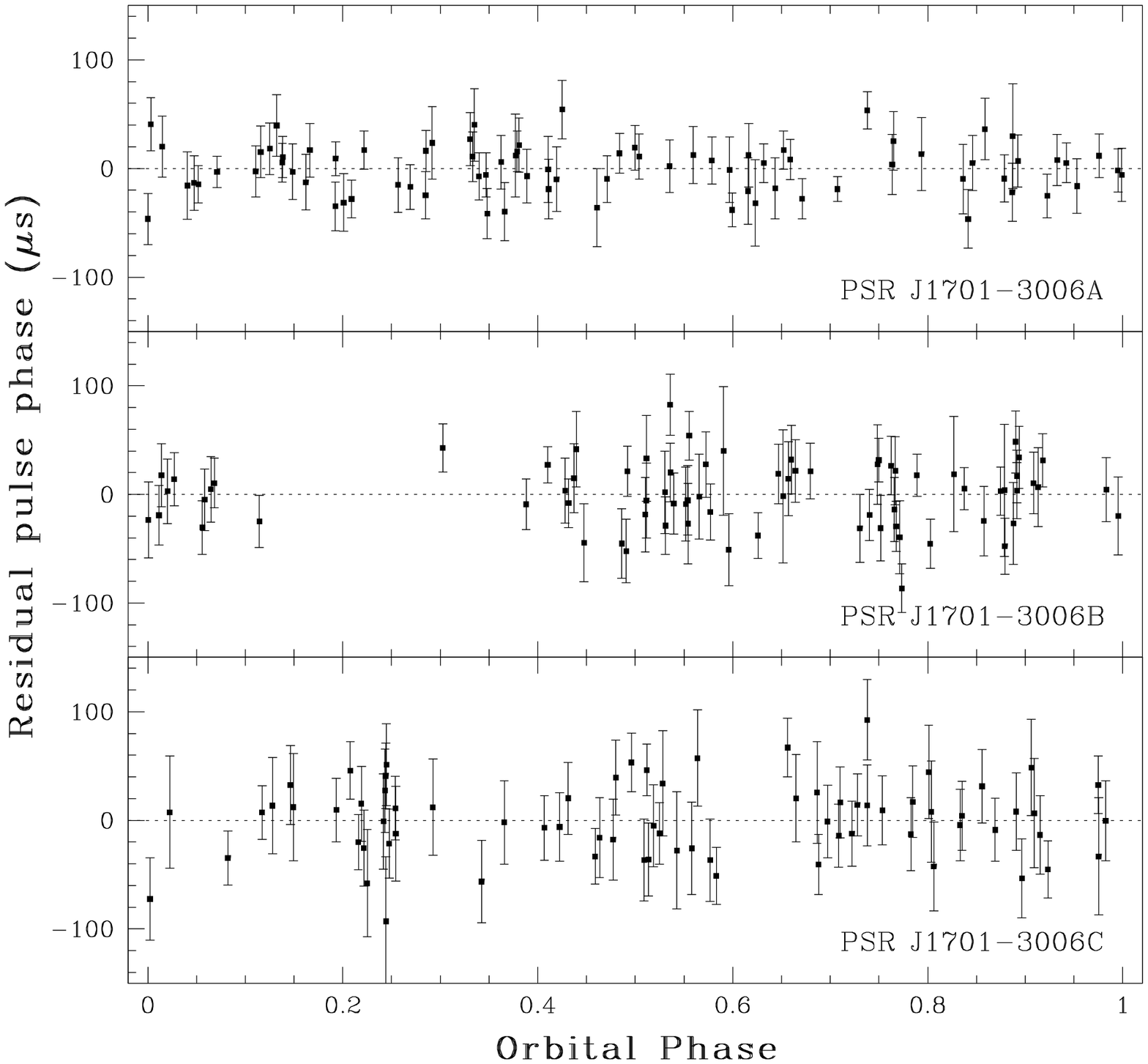}  
\figcaption[possenti2.eps]{\label{fig2}
{Post-fit timing residuals as a function of orbital phase
for the three millisecond pulsars in NGC~6266 discussed in this paper. All
the orbits have been uniformly sampled, with the exception of
PSR~J1701$-$3006B for which we have excluded from the fit the TOAs in
the region of the eclipse.}}
\vskip 0.5truecm}

The inferred radio luminosities of the three millisecond pulsars
($\sim 10-20$ mJy kpc$^2$ at 1400 MHz, corresponding to a luminosity
at 400 MHz L$_{400}\gapp 100$ mJy kpc$^2$ for a typical spectral index
$ -1.7$, see Table~\ref{tab:pars}) places all these sources in the
bright tail of the luminosity function of millisecond pulsars in the
Galactic disk (Lyne et al. 1998\nocite{l+98}) and in 47~Tucanae
(Camilo et. al 2000\nocite{clf+00}).  If we assume a luminosity
distribution ${\rm d}N\propto L^{-1}{\rm d}\log L$ (Lorimer
2001\nocite{l01}), NGC~6266 would contain a few hundred pulsars with
L$_{400}\gapp 1$ mJy kpc$^2$, the approximate limiting luminosity
observed for Galactic disk pulsars. Unfortunately, the
cluster distance and the lack of any strong signal enhancement due
to scintillation will make difficult detecting the fainter pulsar
population, probably preventing a direct investigation of the shape of
the pulsar luminosity function in this cluster.

The pulsar PSR J1701$-$3006A has the largest flux density and the
longest orbital period among the three and was first detected in an
observation during 1999 December (D'Amico et
al. 2001a\nocite{dlm+01}). PSR J1701$-$3006B and J1701$-$3006C are
weaker pulsars in closer binary systems and were confirmed in 2000 November.
Once the orbits were determined, signals from these two pulsars
were recovered in all the observations performed prior to their
confirmation. Therefore the timing solutions reported in
Table~\ref{tab:pars} (and whose residuals are displayed in
Fig.~\ref{fig1}) cover the same time-span for all three MSPs, from
2000 June to 2003 May.  Inspection of Figure~\ref{fig2} shows that
all the orbits have been uniformly sampled (excepting
PSR~J1701$-$3006B for which we have excluded TOAs in the
region of the eclipse; see below) and that there are no systematic
trends in the residuals as a function of binary phase.

\section{Constraints on pulsars and cluster parameters}
\label{3msp}
				   
NGC~6266 is listed in the Webbink (1985\nocite{w85}) catalog as a
moderately reddened, ${\rm E(B-V)}=0.48$, medium-low metallicity
globular cluster, with ${\rm [Fe/H]}=-1.38\pm0.15$, located at $\sim
6.9\pm 1.0$ kpc from the Sun (Brocato et al. 1996\nocite{bbmp96}) and
probably having a collapsed core (Harris 1996\nocite{h96}).

The three millisecond pulsars discussed here are all located close to the
center of mass of the cluster, at least in projection, with projected
distances $\lapp 1.8~\Theta_{c},$ where $\Theta_{c}=10\farcs8$ is the
core radius of NGC~6266 (Harris 1996\nocite{h96}).  This is consistent
with the hypothesis that the cluster has reached thermal equilibrium,
in which energy equipartition gives less velocity to the most massive
species, constraining them to reside deep in the cluster potential well.

The spin period derivatives $\dot{P}$ are all negative, implying that
the line-of-sight acceleration $a_l$ imparted to the pulsars is
directed toward the observer and that it overcomes the (positive)
$\dot{P}_i$ due to intrinsic spin-down (see e.g. Phinney
1993\nocite{phi93}). The probability that a nearby passing star in the
crowded cluster core is significantly accelerating at least one of the
three MSPs is $<1\%$ (Phinney 1993\nocite{phi93}). Moreover, given the
position and the kinematics of the globular cluster NGC~6266, the
centrifugal acceleration of the system (Shklovskii 1970\nocite{s70})
and the vertical acceleration in the Galactic potential (Kuijken \&
Gilmore 1989\nocite{kg89}) produce only negligible effects on the
measured $a_l=|c\dot{P}/P|.$ The differential Galactic rotation
(Damour \& Taylor 1991\nocite{dt91}) can contribute at most a positive
$\sim$10\%, $\sim$2\% and $\sim$25\% to the observed $\dot{P}/P$ of the
pulsars A, B and C respectively; hence, we conclude that the
sign of the line-of-sight accelerations is dominated by the radially symmetric
mean gravitational field of the globular cluster and that the three
MSPs are located behind the plane of the sky through the cluster
center.

The maximum possible $a_l$ due to the mean gravitational field in
NGC~6266 is given by the following relation (accurate at the 10\% level for
$\Theta_\perp \lapp 2 \Theta_c,$ Phinney 1992\nocite{phi92}) 
\begin{equation}
a_{l,{\rm max}}~=~(3/2)~\frac{\sigma_l^2}{D~
(\Theta_c^2+\Theta_\perp^2)^{1/2}}~~,
\label{maxacc6266}
\end{equation}
where $\sigma_l=14.3\pm 0.4$ km s$^{-1}$ is the line-of-sight velocity
dispersion (Dubath et al. 1997\nocite{dmm97}) and $D=6.9\pm 1.0$ kpc
is the distance (Brocato et al. 1996\nocite{bbmp96}). $\Theta_c$ and
$\Theta_\perp$ are the angular core radius and the angular
displacement with respect to the globular cluster center, located at
R.A. (J2000): $17^{\rm h}01^{\rm m}12\fs8$, Dec. (J2000):
$-30^{\circ}06{\arcmin}49{\arcsec}$ (Harris 1996\nocite{h96}, catalog
revision 2003). In particular, for a pulsar with negative $\dot{P}$
the following inequality must hold:
\begin{equation}
\left | \frac{\dot{P}}{P}(\Theta_\perp) \right | = 
\left | \frac{a_{l}}{c}(\Theta_\perp) \right | - \frac{\dot{P}_i}{P} < 
\frac{a_{l,{\rm max}}(\Theta_\perp)}{c}
\label{maxratio6266}
\end{equation}
where $c$ is the speed of light. 

The observed lower limit on the magnitude of the line-of-sight
acceleration of PSR J1701$-$3006B, $a_l=2.9\times 10^{-6}$
cm~s$^{-2}$, is the third largest after those of PSR B2127$+$11A and D
in M15 (Anderson et al. 1990\nocite{agk+90}) and is almost identical
to those of the two MSPs with negative $\dot{P}$ recently discovered
in the central regions of NGC~6752 (D'Amico et
al. 2002\nocite{dpf+02}).  For NGC~6752, the high values of $\dot{P}$
imply a central mass-to-light ratio larger than that from optical
estimates (D'Amico et al. 2002\nocite{dpf+02}). For NGC~6266 on the
other hand, the upper panel in Figure \ref{fig3} shows that the
parameters derived from optical observations can entirely account for
the large $\dot{P}/P$ of PSR J1701$-$3006B (the vertical size of the
dots in Figure \ref{fig3} represents the contribution to $a_l/c$ due to the
differential Galactic rotation).  In particular, applying equation (1)
of D'Amico et al.  (2002\nocite{dpf+02}) we derive a lower limit on
the central mass-to-light ratio (expressed in solar units) for
NGC~6266, ${\cal M}/{\cal L}=1.6$, which is compatible with the
optical value reported in the literature, $2.0$ (Pryor \& Meylan
1993\nocite{pm93}). Similarly, using the observed $\dot{P}/P$ of
J1701$-$3006A (corrected for the Galactic contribution) and equation
(7) of Camilo et al. (2000\nocite{clf+00}) the inferred lower limit
$\rho_0=2.1\times 10^5~{\rm M_\odot pc^{-3}}$ of the central mass
density of NGC~6266 is within the limits obtained from optical data
(Pryor \& Meylan 1993\nocite{pm93}).  These results suggest that, even
though all three clusters display a compact core and very high
line-of-sight accelerations for the embedded pulsars, the dynamics in
the inner region of NGC~6266 probably more resemble those of M15, for
which $2<{\cal M}/{\cal L}<3$ was inferred by Phinney
(1993\nocite{phi93}).

{\vskip 0.5truecm 
\epsfxsize=8.5truecm 
\epsfysize=9.0truecm
\epsfbox{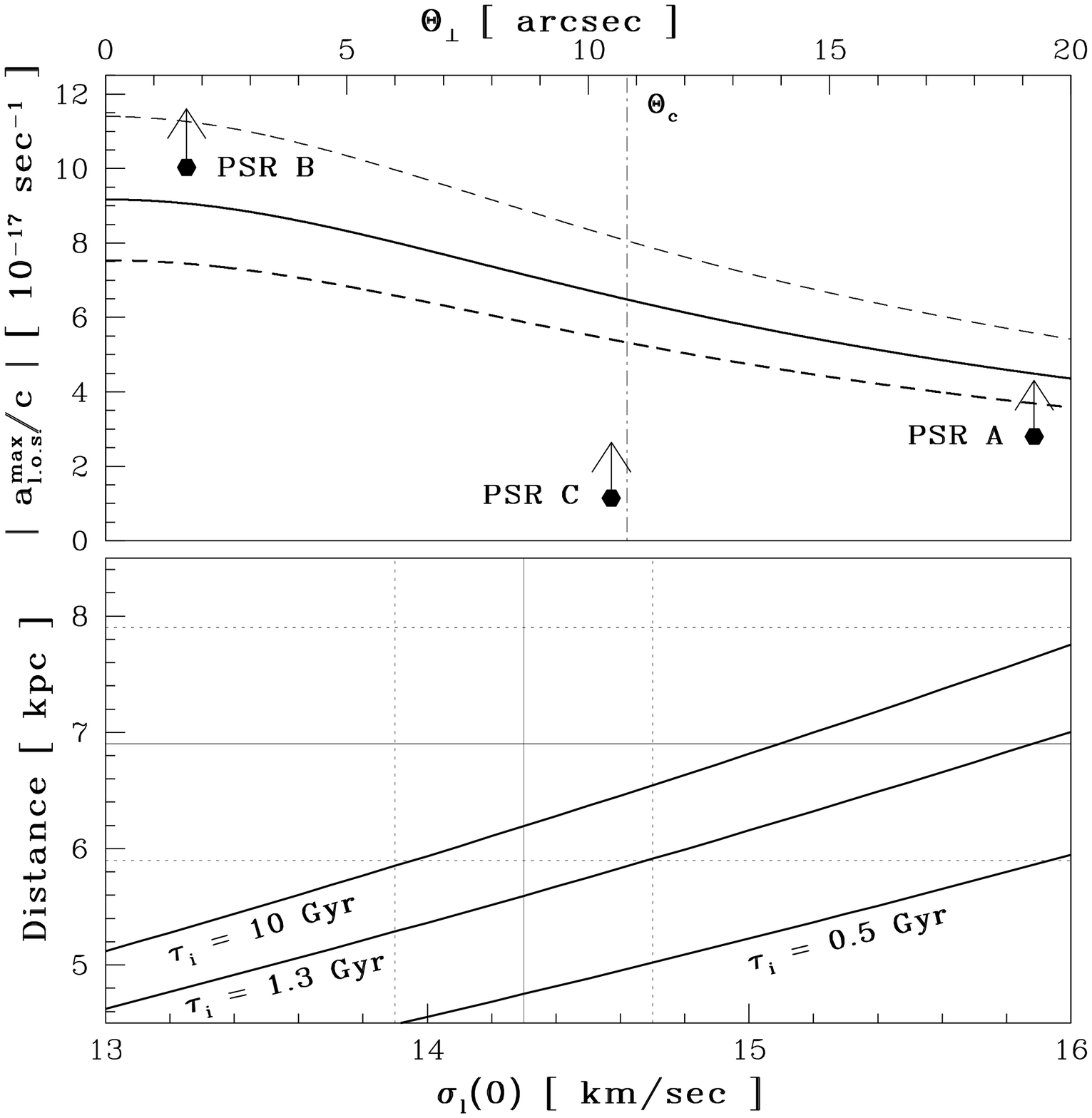} 
\figcaption[possenti3.eps]{\label{fig3}
{{\it Upper panel:} maximum line-of-sight acceleration
$|a_{l,{\rm max}}/c| = |\dot{P}/P|$ versus displacement $\Theta_\perp$
with respect to the center of NGC~6266.  The solid and the two dashed
lines represent the predictions based on equation (\ref{maxacc6266})
using the nominal values of the distance and of the line-of-sight
dispersion velocity and their $1\sigma$ uncertainties obtained from
the available optical observations (see text). The dot-dashed vertical
line marks the assumed angular core radius $\Theta_c=10\farcs8$
(Harris 1996).  The points represent lower limits to the line-of-sight
accelerations based on the timing solutions for the three millisecond
pulsars. The vertical size of the points corresponds to the
contribution to $|\dot{P}/P|$ due to the Galactic potential.  {\it Lower
panel:} constraints on the age of PSR J1701$-$3006B obtained from
equation (\ref{maxacc6266}) and (\ref{maxratio6266}).  The thin solid
lines and the dotted lines represent the values of the parameters
reported in literature and their 1$\sigma$ uncertainties. An
intrinsic characteristic age of PSR J1701$-$3006B larger than about
1 Gyr is compatible with the available observations.}}
\vskip 0.5truecm}

The satisfactory match between the dynamical parameters of NGC~6266
constrained from pulsar timing observations and derived from optical
data allows use of the latter for deriving reliable constraints on the
age and surface magnetic field of the millisecond pulsars. For
instance, the lower panel in Figure~\ref{fig3} shows that the
intrinsic characteristic age of PSR J1701$-$3006B should be greater
than $\sim 1.3$ Gyr to be consistent with the cluster's distance and
velocity dispersion (including their $1\sigma$ uncertainties). This in
turn implies an upper limit on the surface magnetic field
$B_s=3.2\times 10^{19}(P\dot{P})^{1/2}=4.0\times 10^8$ G. Less
stringent limits can be similarly derived for PSR J1701$-$3006A
($\tau_i\gapp 0.15$ Gyr and $B_s\lapp 17\times 10^{8}$ Gauss) and PSR
J1701$-$3006C ($\tau_i\gapp 25$ Myr and $B_s\lapp 31\times 10^{8}$
Gauss). These values are typical for MSPs, both in GCs and in the
Galactic disk.

\section{Range of dispersion measures}
\label{dms}

The MSPs in NGC~6266 show the second largest range in DM (a maximum
deviation $\Delta{\rm DM=0.9}$ cm$^{-3}$pc with respect to the average
${\rm DM_{ave}}=114.34$ cm$^{-3}$pc) after PSR B1744$-$24A and PSR
J1748$-$2446C in Terzan 5 (Lyne at al. 2000\nocite{lmbm00}). This
large range is probably due to a significant gradient in the Galactic
electron column density across different lines-of-sight toward the
cluster, an interpretation supported by the strong variations in
reddening observed across this cluster: $\delta{\rm E}=\Delta{\rm
E(B-V)/E(B-V)}=0.19/0.48$ for an angular displacement of
$\Delta\theta_{\rm E}\sim 7\arcmin$ (as derived from Fig.~3 of
Minniti, Coyne \& Claria 1992\nocite{mcc92}). Alternately, the
variations may have a local origin as in the case of 47~Tucanae, where
pulsar timing observations show that they are due to a plasma
permeating the cluster (Freire et al. 2001\nocite{fklcmd01}). The same
explanation has been proposed in the case of M15 (Freire et
al. 2001\nocite{fklcmd01}) and NGC~6752 (D'Amico et
al. 2002\nocite{dpf+02}). For NGC~6266, the electron number density of
a uniform fully ionized gas would be surprisingly high, $n_e=1.6\pm
0.4$ cm$^{-3}$, at least an order of magnitude larger than that
estimated for the other clusters. The determination of positions,
accelerations and precise DMs of the additional MSPs discovered in this
cluster (Jacoby et al. 2002\nocite{jcb+02}) will help in determining
the origin of the scatter in DM.

\section{The absence of isolated pulsars in NGC~6266}
\label{allbinary}

In contrast to other globular clusters in which at least five pulsars
have been discovered (in order of decreasing number of pulsars:
47~Tuc, M15 and NGC~6752), all the MSPs known in NGC~6266 are in
binary systems (including the three detected by Jacoby et
al. 2002\nocite{jcb+02}).  The absence of known isolated pulsars in
NGC~6266 cannot simply be ascribed to a selection effect since, for a
given spin period and flux density, an isolated MSP is easier to
detect than a binary MSP. Unfortunately, the observational biases
affecting the fraction ${\cal F}_{\rm is}$ of isolated pulsars
discovered in a given cluster (with respect to the total observed MSP
population) are difficult to quantify precisely. Considering all the
other clusters, ${\cal F}_{\rm is}\gapp 2/5$. If this ratio 
applies to NGC~6266, the probability of having the first six detected
pulsars be all binary is $\lapp 5\%$.

If this absence of isolated pulsars in NGC~6266 is not a statistical
fluctuation, it must relate to the mechanisms of formation of these
objects and their interplay with the dynamical state of the cluster.
The few isolated millisecond pulsars observed in the Galactic disk
(where ${\cal F}_{\rm is}\sim 1/3$) are thought to be endpoints of a
rare process of ablation and eventually evaporation of the companion
star by the energetic flux of particles and electromagnetic waves
emitted by the pulsar (e.g. Ruderman, Shaham \& Tavani
1989\nocite{rst89}). Besides this formation channel, the isolated MSPs
seen in globular clusters can also result from close stellar
encounters disrupting a binary system which had previously been
through the recycling process (e.g. Sigurdsson \& Phinney
1993\nocite{sp93}).

This suggests that NGC~6266 is now in a dynamical state where the rate
${\cal R}_{\rm form}$ of formation (and of hardening) of binary
systems containing a neutron star (and suitable for producing new
MSPs) is much larger than the rate of disruption ${\cal R}_{\rm disr}$
of such systems. This idea is supported by comparison of the relevant
rates with other clusters. Table~\ref{tab:ratios} summarizes the
values of ${\cal R}_{\rm form}$ and ${\cal R}_{\rm disr}$ for the four
clusters containing at least five known pulsars. ${\cal R}_{\rm form}$
scales as the rate of close encounters in the cluster, in turn
proportional to $\rho_0^{1.5}r_c^2$, where $\rho_0$ is the central
luminosity density and $r_c$ the core radius of the cluster (Verbunt
2003\nocite{v03}).  Inspection of Table~\ref{tab:ratios} shows that
the expected frequency of close encounters in the core of NGC~6266
\footnote{The absence of pulsars in long period orbital systems
(easily destroyed in dynamical interactions) suggests that close
encounters have been occurring at a significant rate in the central
region of NGC~6266 since a time at least comparable to the cluster
relaxation time, $\sim 1.5$ Gyr at the half-mass radius (Harris
1996\nocite{h96}).}  is 40\% larger than that of M15 and seven times
that of NGC~6752. Although the numbers of known pulsars in these
clusters are similar, the comparison of their radio luminosities (see
discussion in {\S}~\ref{Processing}) indicates that NGC~6266 hosts
many more pulsars than NGC~6752, in accordance with the trend
suggested by the values of ${\cal R}_{\rm form}$.

On the other hand, the probability that a binary, once formed, will
experience a further encounter, which may change or split it
(sometimes creating an isolated millisecond pulsar), scales as ${\cal
R}_{\rm disr}\propto\rho_0^{0.5}r_c^{-1}$ (Verbunt 2003\nocite{v03}).
Hence large values of the ratio ${\cal R}_{\rm form}/{\cal R}_{\rm
disr}\propto \rho_0 r_c^3$ should indicate that more neutron
stars are in binary systems than are isolated. As
Table~\ref{tab:ratios} shows, this prediction roughly conforms with the
numbers for the three globular clusters having
collapsed cores: NGC~6266 (six binary pulsars)
has a ratio ${\cal R}_{\rm form}/{\cal R}_{\rm disr}$ a few times
larger than M15 (one binary and seven single pulsars) and
NGC~6752 (one binary and four single pulsars).

Despite these encouraging agreements, the scaling relations may miss
many factors which could strongly differentiate the pulsar population
in the clusters, e.g., the mass-to-light ratio, the mass function in
the core, the neutron-star retention fraction, the period distribution
of the binary systems and the effects of the collapse of the core.
The last point could be especially relevant, as the only non-collapsed
cluster containing more than five known pulsars, 47 Tucanae, fits with
the predictions based on ${\cal R}_{\rm form},$ but does not satisfy
those related to ${\cal R}_{\rm disr}$ (see Table~\ref{tab:ratios})
--- its binary disruption rate should be less than half that of
NGC~6266, but it hosts several isolated millisecond pulsars, with
a value of ${\cal F}_{\rm is}$ similar to that seen in the Galactic
field, where dynamical encounters are unimportant in the formation
of isolated millisecond pulsars.

Detailed numerical simulations are required to investigate if trapping
of almost all the neutron stars in close binary systems can really occur,
for instance, during the phase immediately preceding the core collapse or its
reversal.

\section{The eclipses in PSR~J1701$-$3006B}
\label{eclipsing}

PSR~J1701$-$3006B displays partial or total eclipses of the radio
signal at $1.4$ GHz near superior conjunction, i.e. at orbital phase
$0.25$ (see Fig.~\ref{fig4}), clearly due to gas streaming off the
companion.  A typical event starts at orbital phases in the range
$0.15-0.20$ and ends at orbital phase $\sim
0.35$, hence sometimes displaying a slight asymmetry with respect to
the expected nominal center of the eclipse at phase 0.25.  At both
eclipse ingress and egress, the pulses usually exhibit excess
propagation delays (see Figs.~\ref{fig4} and \ref{fig5}).  The
eclipse region covers up to 20\% of the entire orbit but, as
illustrated in Figure~\ref{fig4}, unpredictable irregularities affect
both the duration and the appearance of the eclipses.  Sometimes the
pulsation remains barely visible (see e.g. the case of Fig.~4a),
while on other occasions the pulse is totally eclipsed for a large
portion of the event (e.g. the cases of Fig.~4e).  In
a favorable case (Fig.~4b), it has been possible to measure a
slight reduction of the s/n of the pulse (although at the $1\sigma$
level only: see caption of Fig.~\ref{fig5}) as the pulsar signal
crosses the region of interaction with the matter released by the
companion.

{\vskip 0.5truecm 
\epsfxsize=8.5truecm 
\epsfysize=9.0truecm
\epsfbox{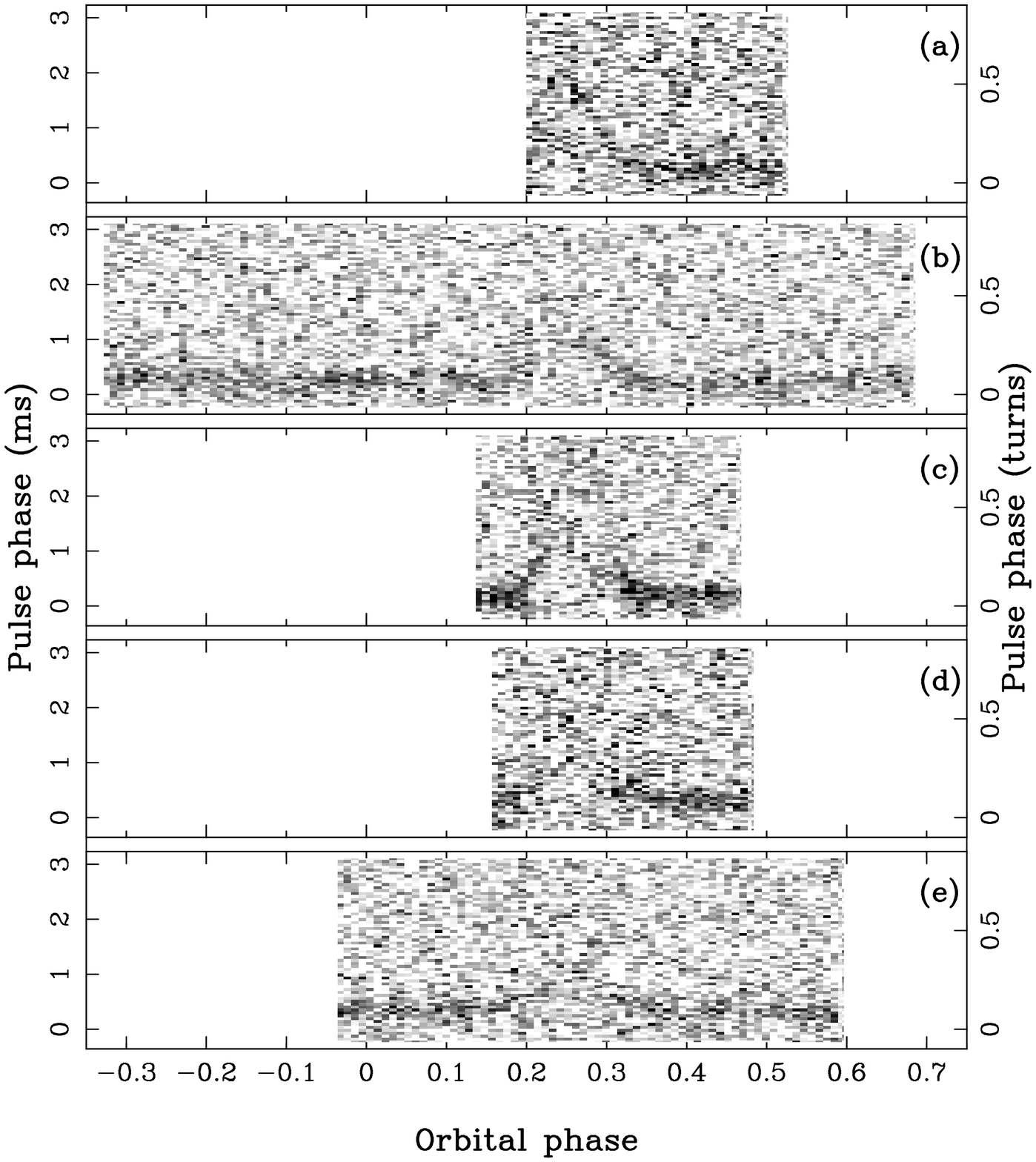} 
\figcaption[possenti4.eps]{\label{fig4}
\footnotesize{Observed signal intensity as a function of orbital phase
and pulsar phase for five observations of PSR J1701$-$3006B centered
at 1390 MHz with a bandwidth of 256 MHz. Eclipses are expected to
occur around superior conjunction (phase 0.25).  The data are
processed in contiguous integrations of 120 s duration.  (a) $\sim 68$
min observation starting on 2002 November 27 at 05:41 UT; (b) $\sim
210$ min observation starting on 2003 January 26 at 00:01 UT; (c)
$\sim 69$ min observation starting on 2000 July 21 at 07:54 UT; (d)
$\sim 68$ min observation starting on 2002 July 10 at 07:12 UT; (e)
$\sim 131$ min observation starting on 2002 April 29 at 13:52 UT.}}
\vskip 0.5truecm}

Pulse broadening and reduction of the s/n prevent investigation of the
frequency-dependent behavior of the delays in our $256-$MHz
bandwidth. However, assuming that they are completely due to
dispersion in an ionized gas (as shown for other eclipsing pulsars,
e.g. Fruchter et al. 1990\nocite{fbb+90}; D'Amico et
al. 2001c\nocite{dpm+01c}), the corresponding electron column density
variations $\Delta N_e$ may represent a first viable explanation of
the eclipse phenomenology. With $\Delta N_e \sim 1.5\times
10^{18}~\Delta t_{-3}$ cm$^{-2},$ where $\Delta t_{-3}$ is the delay
at 1.4 GHz in ms, whenever $\Delta t_{-3}\lapp 2$ (which could be the
case for the entire events in Figs.~4a and 4b), the implied pulse
broadening over the receiver bandwidth $\Delta
P_{-3}=0.36\Delta~t_{-3}$ ms is at most $80\%$ of the intrinsic pulse
width ($\sim 0.50~P$ at 10\% of the peak). Hence the pulse may be only
largely broadened (with an implied reduction of s/n), but not
disappear completely.  On other occasions, the delays may increase
much more rapidly, possibly growing well beyond $\Delta t_{-3}=2.$ In
this case, the DM variations alone could completely smear the signal,
causing a total disappearance of the pulsations.

Alternatively, free-free absorption of the radio-waves in an ionized
envelope of matter released from the companion and expanding adiabatically
can explain both the weakening and the total disappearance
of the radio signal. The optical depth for this process can be written
(see Spitzer 1978\nocite{s78} and Rasio, Shapiro \& Teukolsky
1989\nocite{rsty89}) as
\begin{equation}
\tau_{\rm ff}=0.74\left(\frac{a}{1.32~{\rm R_\odot}}\right)\!
\left(\frac{0.8~{\rm R_\odot}}{R_E}\right)^2\!\!\!\!
\left(\frac{10^4 {\rm K}}{T}\right)^{3/2}\!\!\!\!\!\!
\Delta t^2_{-3}
\label{tau_ff}
\end{equation}
where the orbital separation $a$ and the radius of the eclipse, $R_E$,
defined to be the chord at radius $a$ subtended by the angle between
the orbital phase of eclipse ingress and orbital phase 0.25, are
scaled for PSR J1701$-$3006B (assuming an orbital inclination of 60
degrees, see later), $T$ is the temperature of the fully ionized gas
and $\Delta t_{-3}$ is the observed delay in milliseconds at the
border of the event.  Relatively small delays ($\Delta t_{-3}\lapp
0.4$) imply only a small reduction in the observed flux
density ($\tau_{\rm ff}[\Delta t_{-3}]\lapp 0.1$), whereas $\Delta
t_{-3}\gapp 1$ would be accompanied by significant or complete
absorption of the signal. Interferometric observations of the unpulsed
continuum and observations at other wavelengths will help to clarify
the nature of the eclipses.

{\vskip 0.5truecm 
\epsfxsize=8.5truecm 
\epsfysize=9.0truecm
\epsfbox{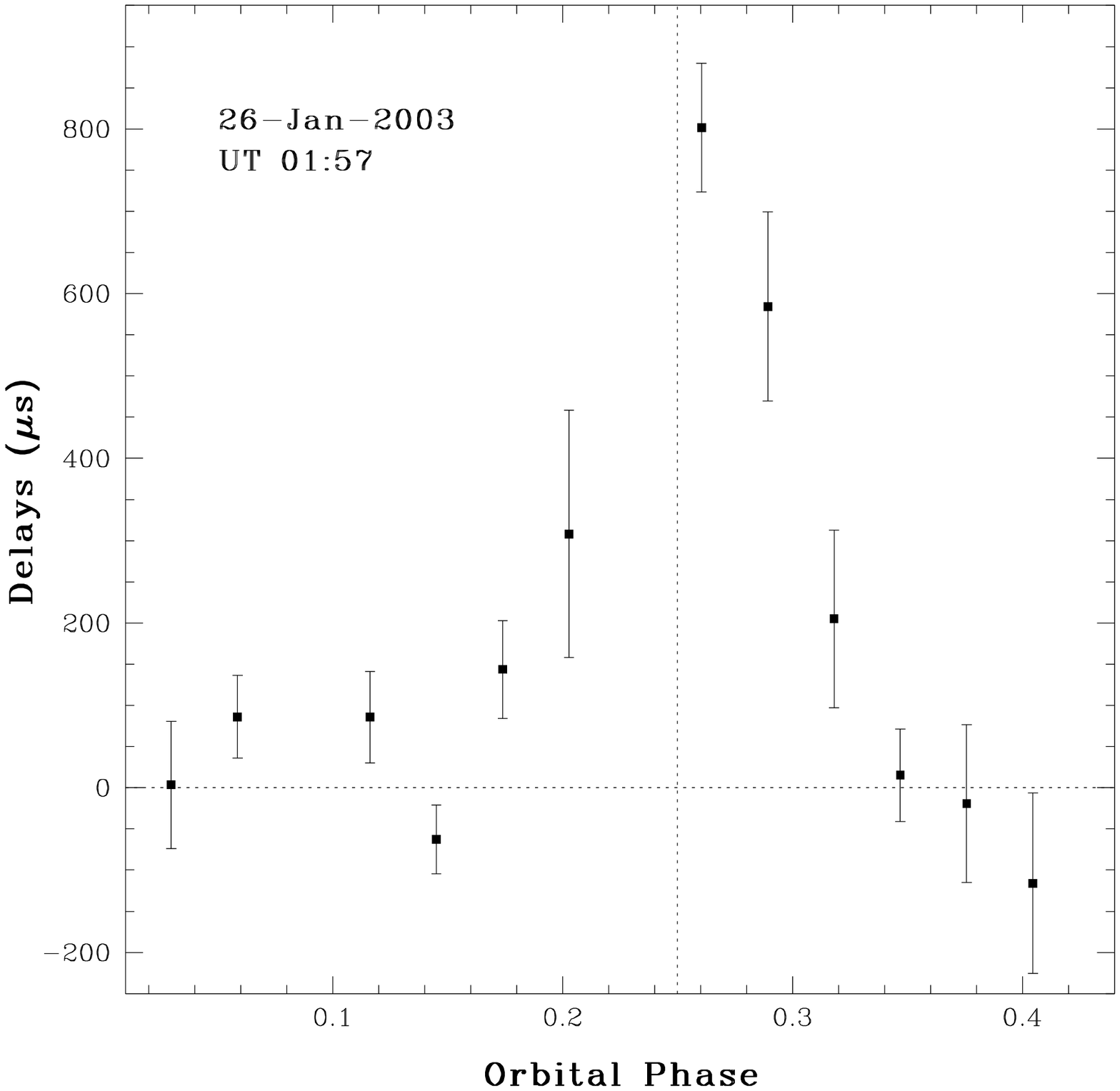} 
\figcaption[possenti5.eps]{\label{fig5}
{Excess group delays of the signal of PSR J1701$-$3006B,
measured on 2003 January 26 (UT time refers to orbital phase
0.25). The observation was centered at 1390 MHz with a bandwidth of
256 MHz and the data are processed in contiguous 360-s integrations.
The error bars are twice the formal uncertainty in the pulse arrival
times. The average value of the s/n within the eclipse
region is $4.6\pm 0.6$, whereas it is $5.7\pm 0.5$ ($1\sigma$
uncertainty) outside.}}
\vskip 0.5truecm}

The occurrence of eclipses suggests that the orbital inclination $i$
is not small. For $i=60^\circ$, the median of all possible inclination
angles, and an assumed pulsar mass of
$1.40~{\rm M_\odot}$, $M_{c,60}=0.14~{\rm
M_\odot}$. For $i\gapp 30^{\circ}$  the companion mass spans the interval
$0.12-0.26~{\rm M_\odot}$, corresponding to a Roche lobe radius in the
range $R_{L}=0.26-0.34~{\rm R_\odot}.$ Hence, independent of the
eclipse mechanism, the extension of the eclipsing cloud, $\gapp
0.8~{\rm R_\odot},$ is larger than $R_{L}$ and the cloud must be
continuously refilled with matter released from the companion.  The
plasma density in the eclipse region is 
$\rho_E\sim 1.6\times 10^{-17} \Delta t_{-3}$ g cm$^{-3}$ and, assuming
isotropic emission, mass continuity implies that the donor star loses
gas at a rate ${\dot M}_{c}=4\pi R_E^2\rho_E v_f\sim
1.0\times 10^{-12}~\Delta t_{-3}~v_{f,8}~{\rm M_\odot yr^{-1}},$ where
$v_{f,8}$ is the wind velocity at $R_E$ in units of 10$^8$
cm s$^{-1}$ (the order of magnitude of the escape velocity from the
surface of the companion).  

If the companion is a helium white dwarf (whose maximum radius is
$R_{wd}=0.04~{\rm R_\odot}$ for masses $>0.12~{\rm M_\odot}$ and
$T\lapp 10^4~{\rm K}$, Driebe et al. 1998\nocite{dsbh98}), and
assuming isotropic emission of the pulsar flux, a significant fraction
$f=(4\% - 20\%)\times(3.7\times 10^{34}{\rm erg~s^{-1}}/\dot{E})$ of
the energy deposited onto the companion surface is necessary for
releasing the observed ${\dot M}_{c}$ (where $\dot{E}$ is the
spin-down power of the pulsar and $3.7\times 10^{34}$ erg s$^{-1}$ its
upper limit derived using the arguments of {\S} 3).
However, the energy requirements are more easily satisfied for a
non-degenerate bloated companion (as appears to be the case in most
eclipsing binary pulsars, Applegate \& Shaham 1994\nocite{as94}). For
example, $f=(0.04\% - 0.2\%)\times(3.7\times 10^{34}{\rm
erg~s^{-1}}/\dot{E})$ for a donor with the radius of a main-sequence
star of the same mass, that is, 3-10 times larger than that of a white
dwarf.  Mass loss from the donor star can be sustained by ablation of
its loosely bound surface layers by the relativistic wind emitted by
the pulsar. This model has been successfully applied to explain
the radio eclipses in close orbital systems having very light
companions, e.g., the cases of PSRs B1957+20 (Fruchter et
al. 1990\nocite{fbb+90}) and J2051$-$0827 (Stappers et
al. 2001\nocite{sbl+01}). As with these other systems, the apparent
mass-loss rate from the companion to PSR J1701$-$3006B is very small;
the ablation time scale $\tau_{\rm abl}=\chi M_{c,60}/{\dot
M}_{c}=\chi 140$ Gyr, where $\chi$ is the ionized fraction, is longer
than the upper limit on the pulsar age (i.e. the cluster age) unless 
$\chi < 0.09$. 

Following an alternate interpretation, the PSR J1701$-$3006B system
may more resemble that of PSR J1740$-$5340, where the effects of the
pulsar irradiation are negligible in triggering the eclipsing wind
from the secondary star (D'Amico et al. 2001c\nocite{dpm+01c}) and the
eclipses (or the excess propagation delays, sometimes seen far away
the nominal phases of eclipse) are caused by matter overflowing the
Roche lobe of the donor star due to the nuclear evolution
of the companion (Ferraro et al. 2001\nocite{fpds01}). In that system,
accretion of matter onto the neutron star is inhibited by the sweeping
effect of the pulsar energetic wind, according to the so-called
radio-ejection mechanism (Burderi, D'Antona \& Burgay
2002\nocite{bdb02}). We note that J1701$-$3006B shares with PSR
J1740$-$5340 {\it (i)} a companion significantly more massive than
those of PSRs B1957+20 and J2051$-$0827, {\it (ii)} the occurrence of
excess propagation delays at 1.4 GHz which are much larger (up to
$\sim 1$ ms vs few tens of $\mu$s) than those observed in any of the
systems having very low mass companions{\footnote{A possible exception
is the pulsar C in the globular cluster M5 (Ransom, private
communication)}} and {\it (iii)} the presence of irregularities in the
eclipses.

A new class of eclipsing recycled pulsars having relatively massive
companions ($M_{c,60}=0.10-0.22~{\rm M_\odot}$) is emerging
from the globular cluster searches. Besides PSR J1701$-$3006B in
NGC~6266 and PSR J1740$-$5340 in NGC~6397, there are PSR B1744$-$24A
in Terzan~5 (Lyne et al. 1990\nocite{lmd+90}), PSR J0024$-$7204W in
47~Tucanae (Camilo et. al. 2000\nocite{clf+00}) and PSR J2140$-$2310A
in M30 (Ransom et. al. 2003a\nocite{r+03}), whereas no similar system
has been detected in the Galactic field to date. A simple explanation
for the overabundance of evaporating ``black widow'' pulsars in
globular clusters with respect to the galactic disk has been recently
proposed by King, Davies \& Beer (2003\nocite{kdb03}): namely, the
current companion of most of the eclipsing pulsars in globulars would
be the swelled descendent of a turn-off star which replaced the
original white dwarf companion of the pulsar in an exchange
interaction in the cluster core. This scenario posits that the
radio-ejection mechanism (Burderi et al. 2001\nocite{b++01}) is now
operating in all the eclipsing millisecond pulsars and provides an
evolutionary basis for separating the systems with very low mass
companion with respect to those having more massive donor stars; in
the former, the mass loss would be driven by angular momentum loss
through gravitational radiation, whereas in the latter the mass loss
rate would be determined by the nuclear evolution of the companion.

The relatively massive systems in globular clusters are good candidates
for optical detection of the donor star{\footnote{In fact, the optical
identification of the secondary star has been recently reported also for
two non eclipsing millisecond pulsars having companions with 
$M_{c,60}\sim~0.2~{\rm M_\odot}:$ PSR J0024$-$7204T in 47 Tucanae
(Edmonds et al. 2003\nocite{e+03}) and PSR 1911$-$5958A in 
NGC6752 (Ferraro et al. 2003\nocite{fpsd03}; Bassa et al. 2003
\nocite{bvvh03})}} and follow-up observations. 
Unlike the Galactic eclipsing systems, their age,
metallicity, extinction, distance and hence intrinsic luminosity and
radius can be estimated from the parent cluster parameters (see, e.g.,
Edmonds et al. 2001a\nocite{eghgc01}; Ferraro et
al. 2001\nocite{fpds01}; Edmonds et al. 2002\nocite{e+02}). In the
case of J1740$-$5340 in NGC~6397, the companion is a red variable star
of magnitude $V \sim 16.5$ (Ferraro et al. 2001\nocite{fpds01}) and
stringent constraints have been set on the effectiveness of the
irradiation of the companion (Orosz \& van Kerkwijk
2002\nocite{ok02}), on the occurrence of the radio-ejection mechanism
(Sabbi et al. 2003\nocite{s++03}) and on the evolutionary path of the
system (e.g. Burderi, D'Antona \& Burgay 2002\nocite{bdb02}; Grindlay
et al.  2002\nocite{g+02}; Ergma \& Sarna 2002\nocite{es03}).

More recently, the companion of the millisecond pulsar J0024$-$7204W
in 47 Tucanae has been optically identified with a blue variable star
of mean magnitude $V\sim 22.3$, probably a heated main sequence star
close to the center of the cluster (Edmonds et
al. 2002\nocite{e+02}). Unfortunately, the pulsar is weak and only
occasionally detectable, which makes the system difficult to
characterize (Camilo et al. 2000\nocite{clf+00}). In the case of PSR
B1744$-$24A in Terzan 5, the strong obscuration toward the Galactic
center ($\gapp 7$ mag in V) prevents detection of the optical
counterpart, even with deep {\it HST} observations (Edmonds et
al. 2001b\nocite{egcl01}).

Consequently, PSR J1701$-$3006B is likely to be a primary candidate
for improving the modeling of eclipsing millisecond pulsars with
relatively massive companions. Indeed, PSR J1701$-$3006B seems to be a
twin of PSR J0024$-$7204W in 47 Tucanae, with similar orbital
parameters and hence minimum companion mass (Camilo et
al. 2000\nocite{clf+00}). Also the pulsar periods are comparable, 3.6
ms versus 2.4 ms. Moreover, unlike PSRs J1740$-$5340 and B1744$-$24A,
both PSRs J1701$-$3006B and J0024$-$7204W reside well within one core
radius of the parent cluster center and hence are in more similar
environments. Assuming that the companion to PSR J1701$-$3006B has the
same luminosity and colors as the companion to PSR J0024$-$7204W, its
photometry would be feasible with deep exposures reaching V-magnitude
24.5. Photometry would of course be much easier if the companion fills
its Roche lobe as is believed to be the case in PSR J1740$-$5340.

The X-ray counterparts of two of the five eclipsing MSPs with
relatively massive companions (namely PSRs J1740$-$5340 and
J0024$-$7204W) have been identified using {\it Chandra} observations
(Grindlay et al. 2001a\nocite{ghem01}, 2001b\nocite{ghemc01}).  Their
spectra appear significantly harder than those of most other known
X-ray counterparts to MSPs in globular clusters. That suggests
(Edmonds et al. 2002\nocite{e+02}) that a non-thermal contribution to
the X-ray emission, perhaps arising from shock interactions at the
interface between the companion and pulsar winds, dominates over the
thermal component seen in the other MSPs, which probably originates
from heated magnetic polar caps on the neutron star. The
identification of the X-ray counterpart of PSR J1701$-$3006B and a
comparison of its X-ray hardness ratio with that of the other MSPs in
NGC~6266 would test the above picture.  Interestingly, 
a long {\it Chandra} pointing towards NGC~6266 shows
that it hosts the largest number of X-ray sources (with luminosity
$>4\times 10^{30}$ erg s$^{-1}$ in the 0.5-6.0 keV range) observed so
far in a globular cluster (Pooley et al. 2003\nocite{p03}).  Possibly
among the 51 detected sources is a significant population of neutron
stars, of which the 6 pulsars discovered so far are a manifestation.

\section{Conclusion}

We have presented rotational and astrometric parameters of three
binary millisecond pulsars located within 1.8 core radii of the
center of the globular cluster NGC~6266. One of these systems, PSR
J1701$-$3006B, displays eclipses for $\sim 20\%$ of the orbit. 
In summary, we note that:
\begin{itemize}
\item[1.] The derived lower limits on the central mass-density
($2.1\times 10^5~{\rm M_\odot}$pc$^{-3}$) and the central mass-to-light ratio
(${\cal M}/$${\cal L}$$\; > 1.6$ in solar units) of NGC~6266 are consistent 
with optical estimates.
\item[2.] The large spread in the dispersion measures of the three
millisecond pulsars is probably due to a significant gradient in the
Galactic electron column density across different lines-of-sight
toward the cluster. 
\item[3.] Even though the nature of the eclipses cannot yet be fully
constrained, the relatively low mass-loss rate from the secondary
star makes unlikely that PSR J1701$-$3006B will evaporate its
companion.
\item[4.] The lack of known isolated pulsars in NGC~6266 is unlikely
to be due to chance or observational bias and suggests that
the cluster is in a dynamical phase favoring formation over
disruption of binary systems containing a neutron star.
\end{itemize}

\acknowledgements {\small{We thank the referee Scott Ransom for very
helpful criticisms of the original manuscript.  ND'A and AP received
financial support from the Italian Space Agency (ASI) and the Italian
Minister of Research (MIUR) under {\it Cofin 2001} national program.
FC acknowledges support from NSF grant AST-02-05853.  The Parkes radio
telescope is part of the Australia Telescope which is funded by the
Commonwealth of Australia for operation as a National Facility managed
by CSIRO.}}

\newpage

\input{tab1}
\input{tab2}


\end{document}

%% file: tab1.tex
\begin{deluxetable}{llll}
\tablewidth{17.5truecm}
\tabcolsep 0.05truecm
\footnotesize
\tablecaption{\label{tab:pars}Observed and derived parameters 
for three pulsars in NGC~6266}
\tablecolumns{4}
\tablehead{
{Pulsar name}                 &
\colhead{PSR~J1701$-$3006A}   &
\colhead{PSR~J1701$-$3006B}   &
\colhead{PSR~J1701$-$3006C}  
}
\startdata
R. A. (J2000)\dotfill               & 17$^{\rm h}$ 01$^{\rm m}$ 12\fs5127(3)  
                                    & 17$^{\rm h}$ 01$^{\rm m}$ 12\fs6704(4)
                                    & 17$^{\rm h}$ 01$^{\rm m}$ 12\fs8671(4)
                                    \\
Decl. (J2000)\dotfill               & $-30\arcdeg$ 06\arcmin 30\farcs13(3)
                                    & $-30\arcdeg$ 06\arcmin 49\farcs04(4) 
                                    & $-30\arcdeg$ 06\arcmin 59\farcs44(4)
                                    \\
Period, $P$ (ms)\dotfill            & 5.2415662378289(16)
                                    & 3.5938522173305(14)
                                    & 3.8064243637728(18)               
                                    \\
Period derivative, $\dot P$\dotfill & $-1.3196(9)\times 10^{-19}$
                                    & $-3.4978(7)\times 10^{-19}$
                                    & $-3.189(11) \times 10^{-20}$ 
                                    \\
Epoch (MJD)\dotfill                 & 52050.0 
                                    & 52050.0
                                    & 52050.0
                                    \\
Dispersion measure, DM (cm$^{-3}$\,pc)\dotfill & 115.03(4) 
                                               & 113.44(4) 
                                               & 114.56(7) 
                                               \\
Orbital period, $P_b$ (days)\dotfill & 3.805948407(16) 
                                     & 0.1445454304(3) 
                                     & 0.2150000713(15)
                                     \\
Projected semi-major axis, $x$ (l-s)\dotfill   & 3.483724(8)
                                               & 0.252775(13)
                                               & 0.192880(12)
                                               \\
Eccentricity\tablenotemark{a}, $e$\dotfill     & $< 4\times 10^{-6}$
                                               & $< 7\times 10^{-5}$
                                               & $< 6\times 10^{-5}$
                                               \\
Time of ascending node, $T_{\rm asc}$ (MJD)\dotfill 
                                               & 52048.5627980(15)
                                               & 52047.2581994(9)
                                               & 52049.855654(2)
                                               \\
Span of timing data (MJD)\dotfill & 51714--52773
                                  & 51714--52773 
                                  & 51714--52773 
                                  \\
Number of TOAs\dotfill            & 80
                                  & 74 
                                  & 73   
                                  \\ 
Rms timing residual ($\mu$s)\dotfill  
                                  & 21   
                                  & 26
                                  & 32
                                  \\
Flux density at 1400\,MHz, $S_{1400}$ (mJy)\dotfill 
                                  & 0.4(1) 
                                  & 0.3(1) 
                                  & 0.3(1) 
                                  \\
\cutinhead{Derived parameters\tablenotemark{b}}
Galactic longitude, $l$ ($\deg$)\dotfill 
                                  & 353.577 
                                  & 353.573
                                  & 353.572
                                  \\
Galactic latitude, $b$ ($\deg$)\dotfill  
                                  & 7.322   
                                  & 7.319  
                                  & 7.316  
                                  \\
Mass function, $f_p$ (M$_{\odot}$)\dotfill
                                                    & 0.00313392(2)
                                                    & 0.00082999(13)
                                                    & 0.00016667(3)
                                                    \\
Companion mass, $M_c$ (M$_{\odot}$)\dotfill
                                                    & $>0.20$ 
                                                    & $>0.12$ 
                                                    & $>0.07$ 
                                                    \\
Radio luminosity, $L_{1400}$ (mJy\,kpc$^2$)\dotfill
                                                    & 19(7)      
                                                    & 14(6)   
                                                    & 14(6)     
                                                    \\
Offset, $\Theta_{\perp}$ ($\arcsec$)\dotfill
                                                    & 19.2 
                                                    &  1.7    
                                                    & 10.5    
                                                    \\
\enddata

\tablecomments{Figures in parentheses are twice the nominal {\sc tempo}
uncertainties in the least-significant digits quoted.}

\tablenotetext{a}{The $2\sigma$ upper limits on the orbital
eccentricities were obtained using the {\sc tempo} ELL1 model, where
$T_{\rm asc}$ and $(e \cos \omega, e \sin \omega)$ are fitted
(Lange et al. 2001\nocite{lcw+01}).  The value given for PSR
J1701-3006B is tentative as not all of the orbit is sampled.  All the
other parameters are derived using the standard (BT) binary model with $e=0$.}

\tablenotetext{b}{The following formulae are used to derive parameters
in the table: $f_p = x^3 (2\pi/P_b)^2 T_\odot^{-1} = (M_c\sin
i)^3/(M_p+M_c)^2$, where $T_\odot \equiv G M_\odot/c^3 = 4.925\,\mu$s,
$M_p$ and $M_c$ are the pulsar and companion masses, respectively, and
$i$ is the orbital inclination angle.  $M_c$ is obtained from the mass
function, with $M_p = 1.40$\,M$_{\odot}$ (\cite{tc99}) and
$i<90\arcdeg$.  The assumed distance is that of the globular cluster,
$d=6.9$ kpc, and $L_{1400} \equiv S_{1400} d^2$. $\Theta_{\perp}$ is the
angular separation in the plane of the sky between the MSP and the
center of NGC~6266 (Harris 1996, revision 2003).}
\end{deluxetable}

%% file: tab2.tex
\begin{deluxetable}{lccllc}
\tablewidth{13truecm}
\footnotesize
\tablecaption{\label{tab:ratios}Encounter and disruption rates for binaries
in four globular clusters}
\tablecolumns{6}
\tablehead{
{Cluster}       &
\colhead{Isolated PSRs}   &
\colhead{Binary PSRs}   &
\colhead{${\cal R}_{\rm form}$}  & 
\colhead{${\cal R}_{\rm disr}$}  &
\colhead{${\cal R}_{\rm form}/{\cal R}_{\rm disr}$}}
\startdata
NGC~6266   &  0  &  6  &  1.4  &  2.5 &  0.57  \\
NGC~6752   &  4  &  1  &  0.19 &  3.4 &  0.056 \\
M~15       &  7  &  1  &  1.0  &  5.5 &  0.19  \\
47~Tucanae &  7  & 15  &  1.0  &  1.0 &  1.00  \\
\enddata
\tablecomments{${\cal R}_{\rm form}$ is estimated as $\propto 
\rho_0^{1.5}r_c^2,$ whereas ${\cal
R}_{\rm disr}\propto\rho_0^{0.5}r_c^{-1}$ (see text for details).
All the values are
normalized to the parameters of 47 Tucanae. Central luminosity density
$\rho_0$ and core radius $r_c$ are obtained from the catalog of
Harris (1996, revision 2003).}
\end{deluxetable}